# Correlations of the first and second derivatives of atmospheric $CO_2$ with global surface temperature and the El Niño-Southern Oscillation respectively

L.M.W. Leggett [1] and D.A. Ball [1]

[1] Global Risk Policy Group Pty Ltd  www.globalriskprogress.com
Correspondence should be addressed to L.M.W.L.
(mleggett@globalriskprogress.com)

*ABSTRACT. Understanding current global climate requires an understanding of trends both in Earth's atmospheric temperature and the El Niño–Southern Oscillation (ENSO), a characteristic large-scale distribution of warm water in the tropical Pacific Ocean and  the dominant global mode of year-to-year climate variability (Holbrook et al. 2009).  However, despite much effort, the average projection of current climate models has become statistically significantly different from the 21$^{st}$ century global surface temperature trend (Fyfe 2013), and has failed to reflect the statistically significant evidence that annual-mean global temperature has not risen in the twenty-first century (Fyfe 2013;  Kosaka 2013). Modelling also provides a wide range of predictions for future ENSO variability, some showing an increase, others a decrease and some no change (Guilyardi et al 2012; Bellenger 2013). Here we present correlations which include the current era and do not have these drawbacks. The correlations arise as follows. First, it has been shown (Kuo 1990, Wang W. et al. 2013) that the rate of change of the level of atmospheric $CO_2$ (expressed as its first derivative) has a statistically significantly similar time-trend signature to that for global surface temperature. Second, we show here that the rate of this change - the second derivative of the level of atmospheric $CO_2$ - is statistically significantly correlated with the separate signature displayed by the El Niño-Southern Oscillation. Third, we show that second-derivative atmospheric $CO_2$ leads ENSO, first-derivative*

*$CO_2$ and temperature. Taken together the foregoing three points provide further lines of evidence for the role of atmospheric $CO_2$ as a key driver of global climate. The results may also contribute to more accurate prediction of future global climate.*

The El Niño Southern Oscillation (ENSO) is a large-scale oceanic warming in the tropical Pacific Ocean that occurs every few years (Wang C. 2013). According to Wang C. (2013): "The Southern Oscillation is characterised by an interannual seesaw in tropical sea level pressure between the western and eastern Pacific, consisting of a weakening and strengthening of the easterly trade winds over the tropical Pacific. … For many decades, it has been recognised that there is a close connection between El Niño/La Niña and the Southern Oscillation, and that they are two different aspects of the same phenomenon…"

Despite much research (IPCC 2007; Bellenger 2013), the mechanisms of ENSO are still not fully understood. In particular, it is not clear how ENSO changes with, or fully interacts with, a changing climate (Guilyardi 2012; Bellenger 2013)

Many causal mechanisms have been proposed for ENSO (Wang C. 2013), including, for example, that it is a self-sustained and natural oscillatory mode of the coupled ocean-atmosphere system, or a stable mode triggered by stochastic forcing.

The issue concerning global surface temperature is that despite the continued rise in greenhouse gas emissions in the 21$^{st}$ century, the trend in global surface temperature has slowed compared to both its previous trend and to climate models. The result is such that first, as mentioned above, the annual-mean global temperature has not risen in the twenty-first century (Kosaka 2013), a result which is statistically significant (Fyfe 2013). Second, this outcome differs from the predictions of mainstream models. Fyfe (2013) considered projected trends in global mean surface temperature computed from a comprehensive set of 117 simulations of the climate by 37 mainstream climate models and determined an average simulated trend. He found that for temperature trends computed over the past fifteen years (1998–2012) the observed trend per decade was more than four times smaller than, and statistically significantly different from, the average trend projected by the models. Fewer than five per cent of the models fitted the observed temperature trend.

The situation is illustrated visually in Figure 1a which shows the increasing departure over recent years of the global surface temperature trend from that projected by a representative climate model (the CMIP3, SRESA1B scenario model for global surface temperature (KNMI 2013)).

In what follows, the following approach is employed (See Methods for fuller information).

Firstly, from the second figure (Figure 1b) to the penultimate figure (Figure 5a) and associated text the tropical surface temperature trend is used for analysis in preference to the overall global average surface temperature trend. This is because the rationale of the investigation is to seek evidence of one clearly depicted phenomenon linking to another. Because of the reverse seasonality between the hemispheres, it is known that using global averages in climate studies can reduce the clarity of temperature phenomena, even to the extent that different effects can approach cancelling each other out. The tropical surface temperature is the average temperature between latitudes 30 degrees North and 30 degrees South. In the final figure (Figure 5b) and associated text the results derived for tropical surface temperature are compared again with global surface temperature.

For the same reason of clarity and simplicity ENSO is depicted just by its Southern Oscillation component, the Southern Oscillation Index (SOI). The SOI takes into account only sea-level pressure. In contrast, the El Niño component of ENSO is specified in terms of changes in the Pacific Ocean sea surface temperature relative to the average temperature. It is considered to be simpler to conduct an analysis in which the temperature is an outcome (dependent variable) without also having (Pacific Ocean) temperature as an input (independent variable). The correlation between SOI and the other ENSO indices is high, so we believe this assumption is robust.

Finally, in the analyses and in the figures, all data are monthly, and are standardly (for example, (Lean and Rind 2008) normalised to a zero mean and variance of 1 for the various periods covered. Again standardly (IPCC 2007), data are subject to various levels of smoothing (low pass filtering) to maximise comparability between series. In general, series with higher levels of differencing require more smoothing. Smoothing is carried out by means of a 13 month moving average (the closeness to a 12 month divisor enables minimisation of seasonal effects, which are not the focus of this study, and the 13 month divisor provides an even number of data points on either side of the mean which enables centring of the resultant smoothed series). If further smoothing is required, a second 13 month smoothing is applied. This is expressed in the text as a 13 month x 13 month moving average.

**Figure 1. Global surface temperature compared to model prediction, first-difference atmospheric $CO_2$, and Southern Oscillation Index (sign reversed to aid comparison)**. **a**) Global surface temperature compared to a IPCC mid-range scenario model (CMIP3, SRESA1B scenario) run for the IPPC third assessment report (IPCC 2007). **b**) Global tropical temperature compared to SOI and first-difference atmospheric $CO_2$. First-difference atmospheric $CO_2$ (blue line) is the Mauna Loa seasonally-adjusted series further smoothed by a 13-month moving average. Sign-reversed SOI (yellow curve) and temperature data (tropical HADCRUT4.2.0.0; purple line) are unsmoothed.

**a**

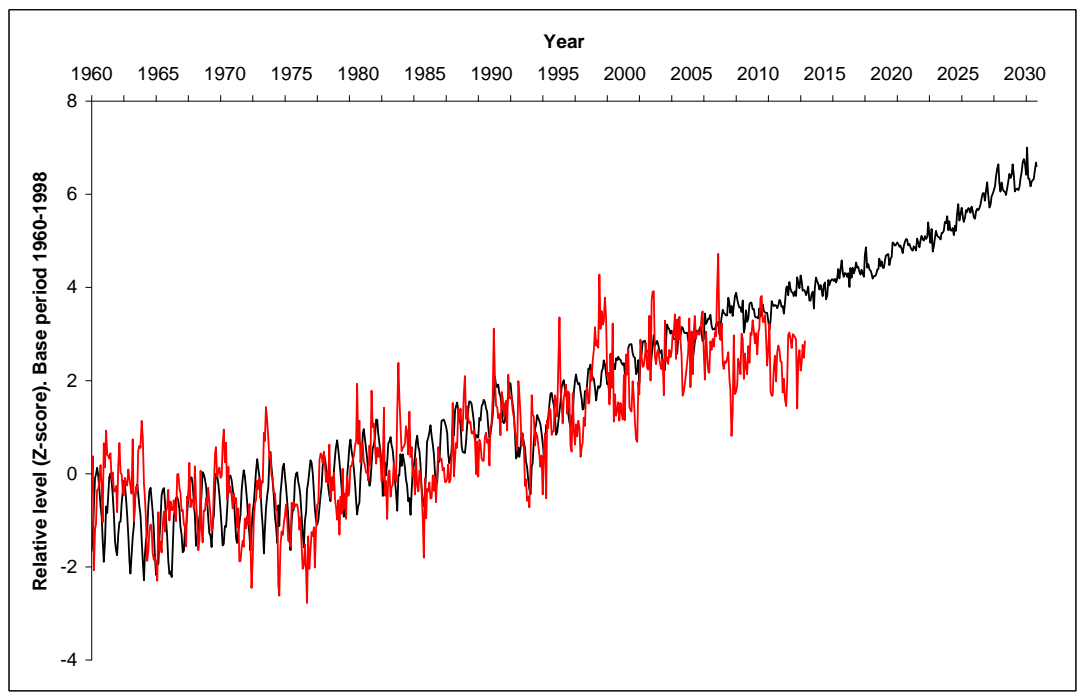

**b**

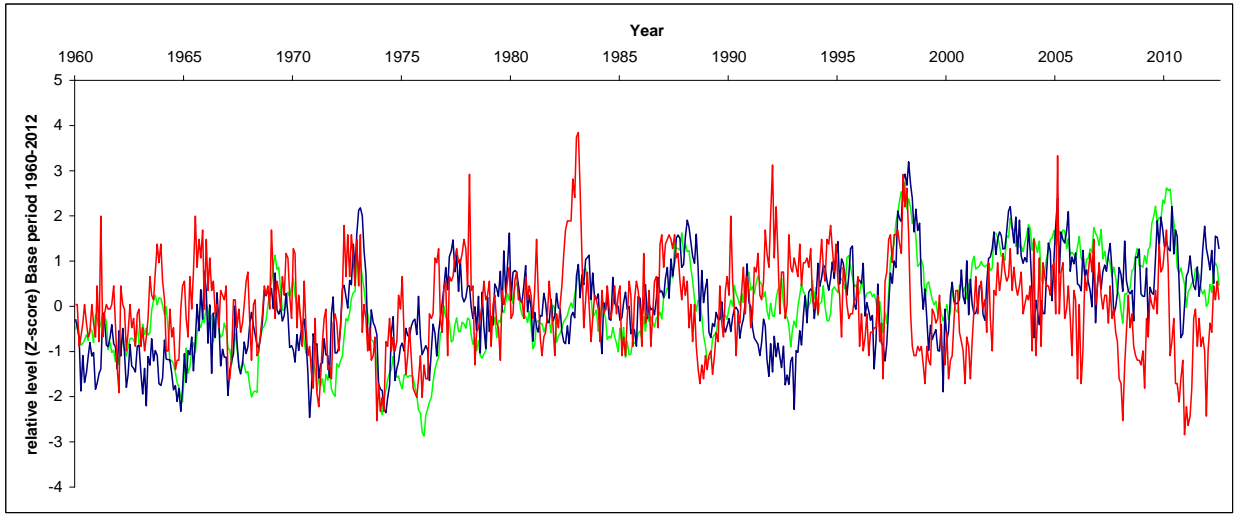

A wide range of causes have recently been proposed to explain the lower-than-expected global surface temperature growth rate since 1998 (Guemas *et al.* 2013). These causes include an increase in ocean heat uptake below the superficial ocean layer; or an effect of the deep prolonged solar minimum, or stratospheric water vapour, or stratospheric and tropospheric aerosols.

Alongside these possible physical causes, Hansen (2013) suggested that the pause in the global temperature increase since 1998 might be caused by the planetary biota, in particular the terrestrial biosphere: that is (IPCC 2007), "the fabric of soils, vegetation and other biological components, the processes that connect them and the carbon, water and energy they store."

It is widely considered that the interannual variability in the growth rate of atmospheric $CO_2$ is a sign of the operation of the influence of the planetary biota. Again, IPCC (2007) states: "The atmospheric $CO_2$ growth rate exhibits large interannual variations. The change in fossil fuel emissions and the estimated variability in net $CO_2$ uptake of the oceans are too small to account for this signal, which must be caused by year-to-year fluctuations in land-atmosphere fluxes."

In the IPCC Fourth Assessment Report, Denman *et al.* (2007) state (italics denote present author emphasis): "Interannual and inter-decadal variability in the growth rate of atmospheric $CO_2$ is dominated by the *response of the land biosphere to climate variations*. …. The terrestrial biosphere *interacts strongly with the climate*, providing both positive and negative feedbacks due to biogeophysical and biogeochemical processes. … Surface climate is determined by the balance of fluxes, which can be changed by radiative (e.g., albedo) or non-radiative (e.g., water cycle related processes) terms. Both radiative and non-radiative terms *are controlled by details of vegetation*."

What in turn might influence the biota's creation of the pattern observed in the trend in the growth rate of atmospheric $CO_2$?

The candidates for the influences on the biota have mainly been considered in prior research to be atmospheric variations, primarily temperature and/or ENSO. Despite its proposed role in global warming overall, $CO_2$ (in terms of the initial state of atmospheric $CO_2$ exploited by plants at time A) has not generally been considered a prime candidate as an influence in the way the biosphere influences the $CO_2$ left in the atmosphere at succeeding time B.

This state of affairs came about for two reasons, one concerning ENSO, the other, temperature. For ENSO, the reason is that the statistical studies are unambiguous that ENSO leads rate of change of $CO_2$ (for example, Lean and Rind, 2008). On the face of it, therefore, this ruled out $CO_2$ as the first mover of the ecosystem processes. For temperature, the reason was that the question of the true phasing between atmospheric temperature and rate of change of $CO_2$ is less settled. Adams (2005): "Climate variations, acting on ecosystems, are believed to be responsible for variation in $CO_2$ increment, but there are major uncertainties in identifying processes (including uncertainty concerning) instantaneous versus lagged responses".

Further, the specific question of the relative effects of rising atmospheric $CO_2$ concentrations and temperature has been addressed by extensive direct experimentation on plants. In a large scale meta-analysis of such experiments, Dieleman et al. (2012) drew together results on how ecosystem productivity and soil processes responded to combined warming and $CO_2$ manipulation, and compared it with those obtained from single factor $CO_2$ and temperature manipulation. While the meta-analysis found that responses to combined $CO_2$ and temperature treatment showed the greatest effect, this was only slightly larger than for the $CO_2$-only treatment. By contrast the effect of the $CO_2$-only treatment was markedly larger than for the warming-only treatment.

The foregoing shows that a strong case can be made for investigating the planetary biota influenced by atmospheric $CO_2$ as a candidate influence on climate outcomes.

This question is explored in the remainder of this paper. The investigation starts by asking, how can the question of a causal influence be addressed from a standard, quantified point of view?

According to Hidalgo and Sekhon (2011), there are four prerequisites to enable an assertion of causality. The first is that the cause must be prior to the effect. The second prerequisite is "constant conjunction" (Hume (1751) cited in Hidalgo and Sekhon (2011)) between variables,. This relates to the degree of fit between variables. The final requirements are those concerning manipulation; and random placement into experimental and control categories. For climate, it is argued that the manipulation criterion is met because the increased $CO_2$ in the atmosphere from human activities is a manipulation. The random assignment criterion is not met but a case can be made that it can be adequately addressed because of the time series nature of the assessments (Gribbons and Herman 1997).

As mentioned, the rate of change of atmospheric $CO_2$ correlates with both the global surface air temperature (for example, see Bacastow 1976; Wang C. *et al.* 2013; Cox *et al.* 2013) and ENSO (for example, Hansen *et al.* 2013). This correlation therefore meets the first-listed of the above prerequisites for a causal relationship. Curves showing the correlations using monthly data from 1960 to 2012 are depicted in Figure 1b.

There is prior research evidence (Raddatz 2007; Wang W. 2013) that the tropical land biosphere dominates the climate–carbon cycle feedback, and therefore the interannual variability of the growth rate of atmospheric $CO_2$. This is confirmed by correlation analysis (Table 1, Supplementary Information). Hence for the research in this study, the tropical (30 degrees N to 30 degrees South latitude) global surface temperature is used as the temperature measure. At the conclusion of the study, the main findings for tropical temperature are compared with those of global temperature and conclusions drawn for global temperature.

In the study, rate of change of atmospheric $CO_2$ is expressed in terms of its derivative. This is done by means of finite differences, which are a convenient approximation to derivatives (Hazewinkel 2001; Kauffman 2006).

Turning to the second-listed causality prerequisite, of degree of fit between variables, a visual inspection of Figure 1b shows that the temporal phasing of the curves varies.

To quantify the degree of difference in phasing between the variables, time-lagged correlations (correlograms) were calculated by shifting the series back and forth relative to each other, one month at a time. Figure 3 shows that the first difference in $CO_2$ is coincident with tropical surface temperature, and the SOI leads both the temperature measure and the first difference in $CO_2$ (by two months). Both relationships are highly statistically significant (see below). These results qualify both the first difference of $CO_2$ and SOI as being considered candidate drivers of temperature. However, as expected from the prior research (see above) SOI also leads the first difference in $CO_2$ (by 2 months). As already discussed this is an initial problem for the notion that the biota rather than SOI might be the cause of the slower-than-expected increase in global temperature since 1998.

Figure 2, however, also displays second-difference $CO_2$ and shows that it leads SOI and temperature.

**Figure 2. Correlograms for the proposed factors that influence global temperature. a)** Temperature as a function of first-difference $CO_2$ (blue curve), as a function of SOI (purple curve), and as a function of second-difference $CO_2$ (yellow curve).

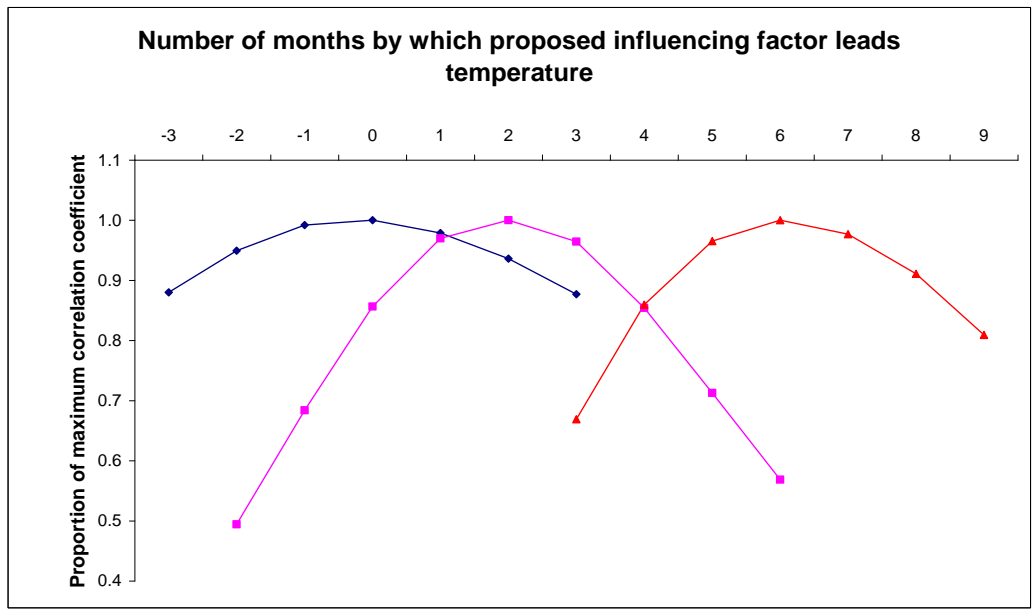

If second-difference $CO_2$ can be entertained as a driver of climate variables, by what mechanism might it act? In contrast to the land biosphere-based first-difference effects described above by Denman (2007), at present a search has found no published information on second-difference sensitivity in plants. Second-difference sensitivity, however, is commonly known in animal sensory systems – for example, in the form of acceleration detectors for limb control (Vidal-Gadea *et al.* 2010). Indeed Spitzer and Sejnowski (1997) argue that rather than occurring rarely, such differencing and other computational processes are potentially ubiquitous in living systems, including at the single-celled level: "Are there principles of information processing common to all biological systems, whether simple or complex, fast or slow? … (There are) many ways in which biochemical reactions within cells can be used for computation. A variety of biological processes — concatenations of chemical amplifiers and switches — can perform computations such as exponentiation, differentiation, and integration."

Plants with the ability to detect the rate of change of scarce resources would have a clear selective advantage. First and second differences, for example, are each leading indicators of change in the availability of a given resource. Leading indicators of change in $CO_2$ would enable a plant's photosynthetic apparatus to be ready in advance to harvest $CO_2$ when, for seasonal or other reasons, increasing amounts of it become available. In this connection, it is noteworthy that second-difference capacity would provide greater advance warning than first.

Has $CO_2$ ever been such a scarce resource? Ziska (2008) states: "…plants evolved at a time of high atmospheric carbon dioxide (4-5 times present values), but concentrations appear to have declined to relatively low values during the last 25-30 million years (Bowes 1996). Therefore, it has been argued ( Körner 2006), for the last c. 20 million years, terrestrial plant evolution has been driven by the optimisation of the use of its scarce 'staple food', $CO_2$.

Given that plants are sensitive to first-difference $CO_2$, and could potentially be sensitive to second-difference $CO_2$, let us turn to seeing the extent to which first- and second-difference $CO_2$ correlates with global surface temperature and other relevant variables.

The method of assessment used is multiple linear regression. This method has frequently been used to quantify the relative importance of natural and anthropogenic influencing factors on climate outcomes such as global surface temperature – for example, Lean and Rind, (2008), Lean and Rind (2009); Foster and Rahmstorf, (2011); Kopp and Lean, (2011); Zhou and Tung, (2013).

From such studies, a common set of main influencing factors (also called explanatory or predictor variables) has emerged. These are (Lockwood (2008); Folland (2013); Zhou and Tung (2013):  El Nino–Southern Oscillation (ENSO), or Southern Oscillation alone (SOI); volcano aerosol optical depth; total solar irradiance; and the anthropogenic warming trend (termed here the four-predictor model). In these models, ENSO/SOI is the factor embodying interannual variation.

Figure 1 in Imbers et al. (2013) shows that a range of different studies using these variables have all produced similar and close fits with the global surface temperature.

In the following section, standardly available versions of the variables used in the following cross-section of such studies [Lean and Rind (2008), Lockwood (2008); Lean and Rind (2009), Kopp and Lean (2011), Wang W. (2013)] are obtained. These are first used to conduct a multiple regression analogous to that of the studies. This multiple regression involves the dependent variable of tropical surface temperature regressed against the predictor variables of level of atmospheric $CO_2$, SOI, volcano aerosol optical depth; and total solar irradiance. Further regression results - both multiple and single, using selected subsets of predictor variables- are also prepared, including now introducing the first-difference $CO_2$ series. Results generated are then compared with the previously published results in Figures 3a and b and Table 1.

Figure 3. **Tropical surface temperature compared to empirical models.** **a**, Observed monthly mean tropical surface temperature (HADCRUT4 dataset) (black) and a multivariate empirical model (red) that combines four different influences: level of atmospheric $CO_2$, SOI, volcanic aerosols, and solar irradiance. **b**, Observed temperature as for **a** and a univariate empirical model comprising first-difference atmospheric $CO_2$.

**a**

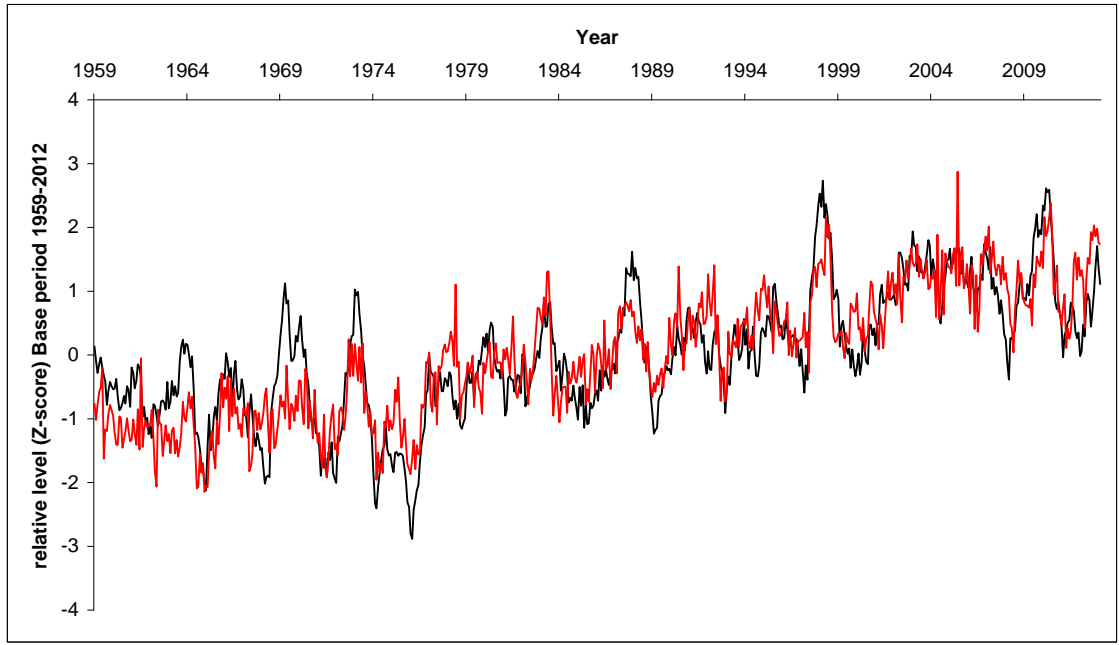

**b**

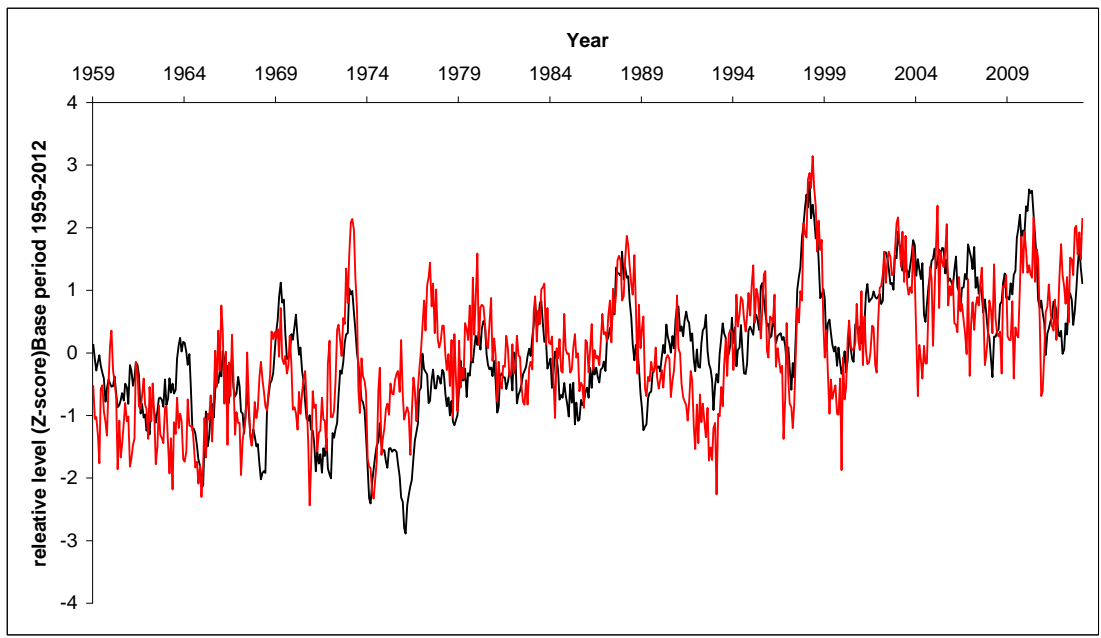

Figure 3a shows the close fit of the presently prepared multiple regression result with global tropical surface temperature. Figure 3b shows the fit of a regression using first-difference $CO_2$ as the only predictor variable. Visual inspection of the figure shows the close fit also achieved.

Analysis of variance statistics for the above two regressions are given alongside equivalent data from the prior studies in the following table.

Table 1: Strength of involvement in predicting tropical surface temperature of potential predictors in nine empirical models. A blank cell signifies data not available.

| Regression Statistics | Full model with soi | Full model with 1st diff $CO_2$ | SOI alone (present paper) | 1st diff alone (present paper) | 1st diff alone (Wang W 2013) | Full model with ENSO (Lockwood 2008) | Full model with ENSO (Kopp and Lean 2011) | Full model with ENSO (Lean and Rind 2009) | Full model with ENSO (Lean and Rind 2008) |
|---|---|---|---|---|---|---|---|---|---|
| Multiple R | 0.82 | 0.79 | 0.37 | 0.7 | 0.7 | 0.89 | 0.92 | 0.87 | 0.87 |
| $CO_2$ Mauna Loa (season corr) | 0.68 | 0.45 | | | | | | | |
| ILed 3m Reverse SOI (KNMI, NCEP) | 0.42 | | | | | | | | |
| 13m ma 1st diff. $CO_2$ Mauna Loa (season corr) | | 0.44 | | | | | | | |
| Led 7m Volc [reverse Global mean aerosol optical depth (incl. proj.)] | 0.14 | -0.02 (ns) | | | | | | | |
| SUN (incl. proj.) | 0.04 (P=0.1) | 0.07 | | | | | | | |

The table shows the following. The standard four-predictor model as run in this study shows a somewhat lower correlation coefficient (multiple R) than the comparison studies, but the result is still classed as a high correlation, and is highly statistically significant.

The multiple regression with first-difference $CO_2$ substituting for ENSO produces a correlation coefficient effectively equal to the SOI-containing multiple regression. Notably, it does this with volcanic aerosols becoming insignificant in the model. This shows that the first-difference $CO_2$ series embodies the effect of the volcanic series. The ENSO alone correlation coefficient is lower, at .37.

The foregoing shows that the first-difference predictor has a role in temperature prediction equivalent to ENSO in the four-predictor model, and as sole predictor produces a much higher correlation than SOI does as sole predictor.

What then, of correlations with second-difference $CO_2$? In the following table, pairwise correlations are provided between all the six variables used in the foregoing regressions plus second-difference $CO_2$, 20 combinations in all. In this analysis, no variables are led or lagged, because the aim is to determine the strength of correlation which arises out of the natural relative phasings.

Table 2: Pairwise correlations of climate variables

|  | Correlation coefficient R |
|---|---|
| Temp by lin. $CO_2$ | 0.708 |
| Temp by 1st diff $CO_2$ | 0.699 |
| Level of $CO_2$ by 1st diff $CO_2$ | 0.57 |
| Volc by 1st diff. $CO_2$ | 0.342 |
| SOI by 2nd diff. $CO_2$ | 0.315 |
| Temp by SOI | 0.31 |
| Reverse SOI by volc | 0.294 |
| Level of $CO_2$ by volc | 0.245 |
| SOI by 1st diff. $CO_2$ | 0.202 |
| Temp by volc | 0.169 |
| Temp by sun | 0.111 |
| Level of $CO_2$ by sun | 0.062 |
| 2nd diff. $CO_2$ by sun | 0.055 |
| 1st diff. $CO_2$ by 2nd diff $CO_2$ | 0.049 |
| 1st diff. $CO_2$ by sun | 0.047 |
| 2nd diff. $CO_2$ by volc | 0.042 |
| SOI by Level of $CO_2$ | 0.025 |
| Volc by sun | 0.023 |
| Temp by 2nd diff $CO_2$ | 0.004 |
| Level of $CO_2$ by 2nd diff $CO_2$ | 0.002 |

Table 2 shows that, out of the 20 correlations depicted, a correlation of second-difference $CO_2$ comes in as the fifth highest. Furthermore, this correlation is with SOI. Even further, this correlation is the highest displayed by SOI with any of the variables.

Let us look (Figure 4a) at the two key pairs of interannually varying factors. For clarity for the purpose of this figure, (i) all curves are given a further 13-month moving average smoothing, and (ii) to facilitate depiction of trajectory, second-difference $CO_2$ and SOI (right axis) are offset so that all four curves display a similar origin in 1960.

Figure 4. **Trends in tropical surface temperature, SOI and first- and second-difference atmospheric** $CO_2$. **a**, Observed monthly mean tropical surface temperature (HADCRUT4 dataset) (black) and first-difference atmospheric $CO_2$ (red)(left-hand scale), and SOI (blue) and second-difference atmospheric $CO_2$ (green) (right-hand scale). **b**, Sign-reversed SOI (unsmoothed and neither led nor lagged) (black); second-difference $CO_2$ smoothed by a 13 month x 13 month moving average and led relative to SOI by 2 months (green); and first-difference tropical surface temperature (HADCRUT4 dataset) smoothed by a 13-month moving average and led by 3 months (red).

**a**

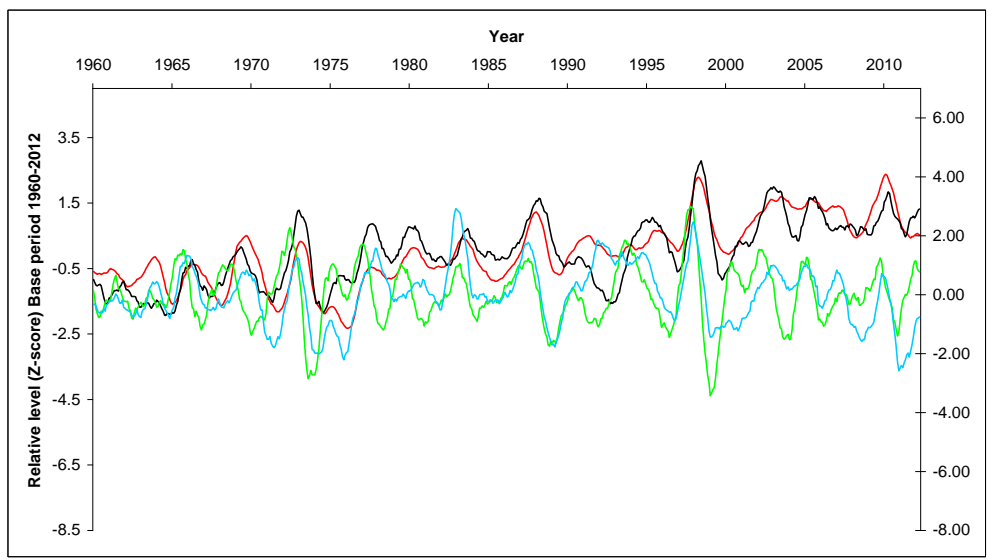

**b**

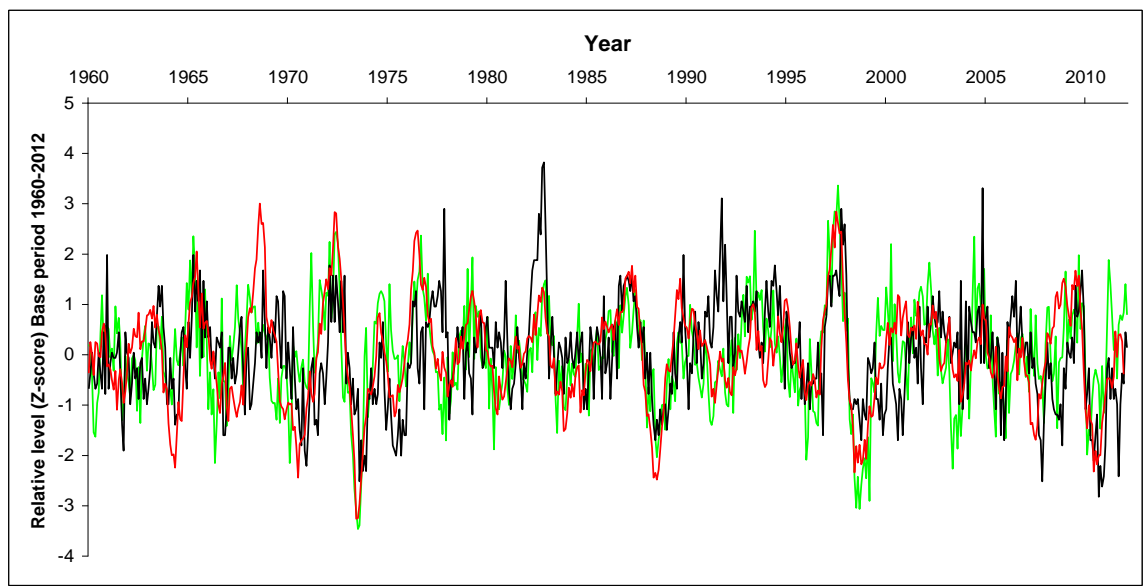

The figure shows that the overall trend, amplitude and phase - the signature - of each pair of curves is both matched within itself and different from the other pair. The remarkable sorting of the four curves into two groups is readily apparent. Each pair of results provides context for the other - and highlights the different nature of the other pair of results.

In the foregoing data analysis, confirmatory analysis has been conducted by means of ordinary least squares. However, as time series are involved, the degree of autocorrelation must be assessed, and if present to a substantial degree, corrected for.

It is noted that autocorrelation does not affect the size of coefficients in the equation linking the independent to dependent variables but does affect the size of correlation coefficient observed and the extent of statistical significance determined.

For the purposes of this study, it is noted that the two main relationships – first difference atmospheric $CO_2$ with tropical surface temperature, and second difference $CO_2$ with SOI - have been assessed for and corrected for autocorrelation. This done, each corrected relationship has been found to be statistically significant (P = .0013 and P = .001 respectively).

The link between all three variable realms — $CO_2$, SOI and temperature — can be further observed in Figure 4b. This shows SOI, second-difference atmospheric $CO_2$, and *first-difference* temperature, each of the latter two series phase-shifted for maximum correlation with SOI (see Table 3). The correlation coefficients for and statistical significance of the correlations between the curves are shown in Table 3. Both results are highly statistically significant.

Table 3. Correlations between SOI, second-difference atmospheric $CO_2$ and first-difference temperature

| Correlation of reverse SOI with: | Correlation coefficient (R) | Significance (P) |
|---|---|---|
| led 2m 13m ma then 13m ma 2nd diff. atmos $CO_2$ (season corr) | 0.36 | 6.88E-21 |
| led 3m 13m ma 1st diff. tropical mean Temp anomaly | 0.46 | 9.77E-35 |

Concerning differences between the curves, two of what major departures there are between the curves are coincide with volcanic aerosols – from the El Chichon volcanic eruption in 1982 and the Pinatubo eruption in 1992 ( Lean and Rind 2009). These factors taken into account, it is notable when expressed in the form of the transformations in Figure 4b that the signatures of all three curves are so essentially similar that it is almost as if all three curves are different versions of - or responses to - the same initial signal. Effectively, temperature is the first derivative of the level of atmospheric $CO_2$, and ENSO the second derivative. Notably, in analogy with kinematics, first-difference $CO_2$ is equivalent to the velocity of change of $CO_2$; and second-difference $CO_2$, the acceleration of change of $CO_2$.

How do these results for the specific category of tropical surface temperature compare with those with global surface temperature overall? Figures 5a and b show trends for each temperature series compared with first-difference $CO_2$.

Figure 5 **Trends in tropical surface temperature and global surface temperature compared with first-difference atmospheric** $CO_2$. **a**, First-difference atmospheric $CO_2$ (black) and observed monthly mean tropical surface temperature (HADCRUT4 dataset)(red). **b**, First-difference atmospheric $CO_2$ (black) and observed monthly mean global surface temperature (HADCRUT4 dataset)(red)

**a**

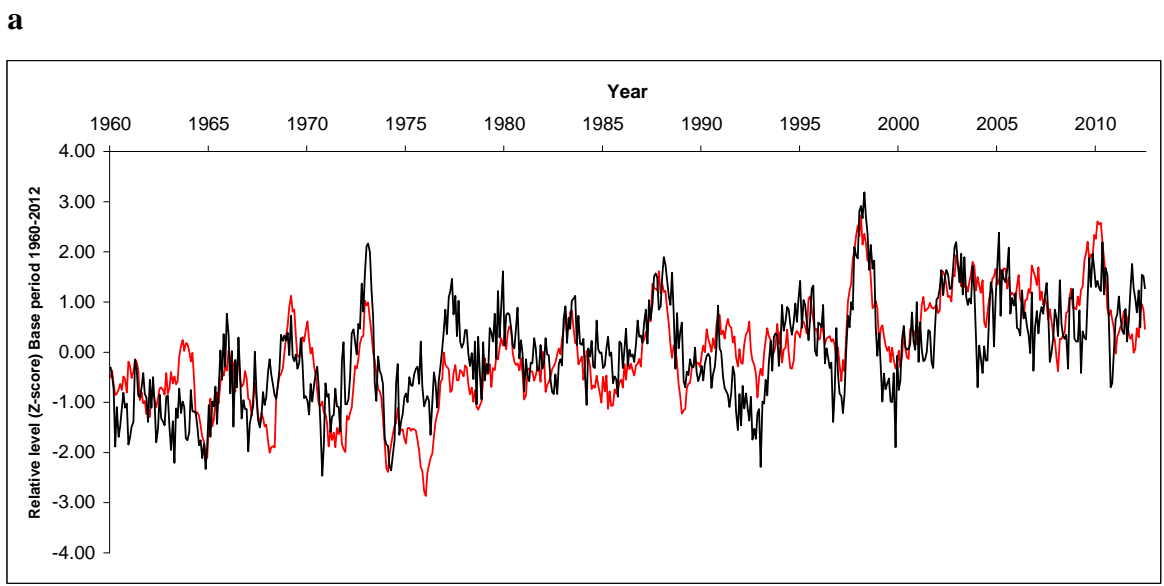

**b**

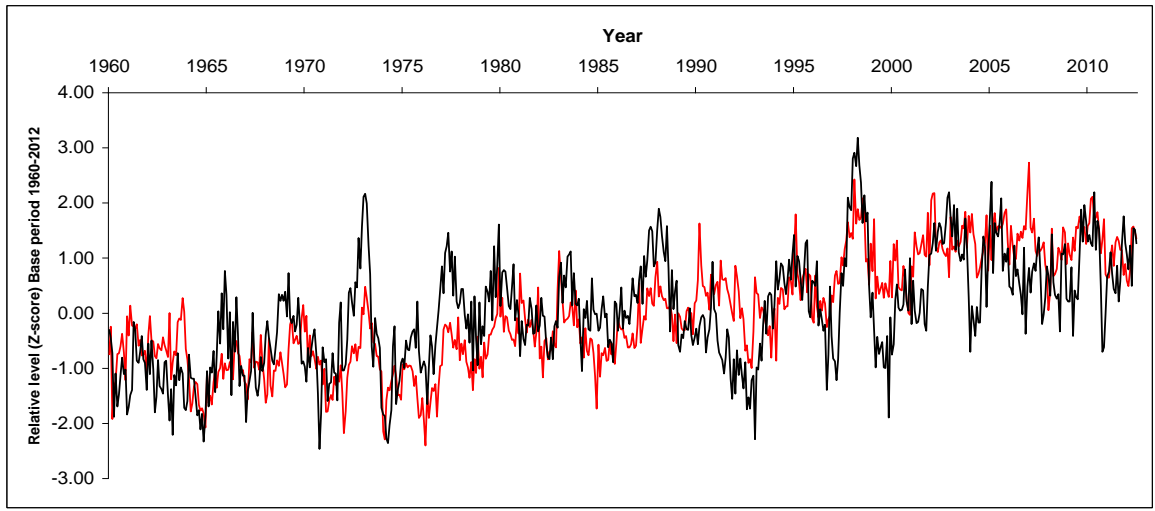

The figures show that each temperature series correlates closely with first-difference atmospheric $CO_2$. It is clear that the tropical correlation is the most precise. This is illustrated by means of correlation analysis: while as already shown (Table 1, models 4 and 5) the correlation coefficient between first-difference $CO_2$ and tropical temperature is 0.70 (P<.0001), that for global temperature is lower, at 0.63 (P<.0001). This higher result for tropical temperature is consistent with the evidence (Raddatz 2007; Wang W. 2013) that the tropical land biosphere dominates the climate–carbon cycle feedback.

As with tropical surface temperature, the correlation of first-difference $CO_2$ with global temperature is strong and highly statistically significant. Hence we consider that the conclusions in the following sections suggested for tropical temperature can also be considered for global temperature.

The preceding results — that the rate of change of global tropical surface temperature reflects first-difference atmospheric $CO_2$, and that SOI reflects second-difference atmospheric $CO_2$ — support several conclusions. First, the extremely close match between SOI and second-difference $CO_2$ is both an explanation for ENSO and strong evidence that climate is sensitive to a second-difference $CO_2$ signal. Second, the close relationship between second- and first-difference $CO_2$ — the latter of which is already widely accepted as originating in the terrestrial biosphere — is strong evidence that the biosphere may drive both ENSO and temperature. Finally, the second-difference $CO_2$/SOI correlation is an indication of the way in which atmospheric $CO_2$ stamps itself on the global climate.

The remarkable closeness of fit between temperature and first-difference $CO_2$ on the one hand, and SOI and second-difference $CO_2$ on the other, suggests a development of the picture of the functioning of the biosphere and the atmosphere and their interrelationships, one in which aspects of the atmosphere are seen as being more closely coupled than previously seen to be (Denman 2007), both to each other and to the biosphere. It may not be too much to suggest that the way aspects of the climate closely follow the different effects may be evidence of the planet's vegetation "conditioning" the atmosphere.

With these results and this perspective, it is interesting that in the very earliest paper showing the correlation between the rate of change of $CO_2$ and ENSO, Bacastow (1976) was open to the possibility that $CO_2$ could drive ENSO: "The possibility exists that the anomaly in the $CO_2$ level drives the Southern Oscillation but it seems unlikely because the $CO_2$ changes are only 1 part per million in a total of about 330 ppm.". The low energy embodied in the rate of change of $CO_2$ that may have concerned Bacastow is of no concern, however, if the rate of change of $CO_2$ is not energy to the plants but information.

If such information then drives the biosphere, what is the scale of the energy it embodies relative to the atmosphere? One estimate of the current average rate of global energy capture by photosynthesis is approximately 130 TW (Steger *et al.* 2005). This is of the order of 10-20 per cent of the power that is associated with all winds within the global atmosphere, estimated at between 900 and 1700 TW (Miller *et al.* (2011)). Another measure of the scale of activity of the terrestrial biosphere relative to the atmosphere is that the loss of water through the stomata of plants' leaves (that is, transpiration) is the single largest mechanism by which all soil moisture is returned to the atmosphere (Schlesinger and Bernhardt 2013).

These relative scales of energy add to the plausibility that changes in plant activity - responding to the information regarding the rate of change of atmospheric $CO_2$ for the purposes of photosynthesis – are massive, and definitely sufficient to substantially modulate climate variables.

With this in mind, concerning ENSO, rather than "a self-sustained and naturally oscillatory mode of the coupled ocean-atmosphere system or a stable mode triggered by stochastic forcing" (Wang C. 2013), it may seem much more plausible for ENSO to be a product of the massive energy flows of the biosphere.

Already proposed candidate causes for the current trend in global surface temperature include (Guemas *et al.* 2013) an increase in ocean heat uptake below the superficial ocean layer; or an effect of the deep prolonged solar minimum, or stratospheric water vapour, or stratospheric and tropospheric aerosols. The present results show that, over the half century period of monthly data studied, the Earth's average global surface temperature correlates very closely throughout with the rate of atmospheric $CO_2$ growth. This is evidence that the rate of atmospheric $CO_2$ growth, too, is a candidate cause.

**Methods**

We used the Hadley Centre–Climate Research Unit combined land SAT and SST (HadCRUT) version 4.2.0.0 http://www.metoffice.gov.uk/hadobs/hadcrut4/data/download.html, the U.S. Department of Commerce National Oceanic & Atmospheric Administration Earth System Research Laboratory Global Monitoring Division Mauna Loa, Hawaii monthly $CO_2$ series (annual seasonal cycle removed) ftp://ftp.cmdl.noaa.gov/ccg/$CO_2$/trends/$CO_2$_mm_mlo.txt, for volcanic aerosols the National Aeronautic and Space Administration Goddard Institute for Space Studies stratospheric aerosol optical depth http://data.giss.nasa.gov/modelforce/strataer/, SOI (Southern Oscillation Index) from National Centers for Environmental Protection www.cpc.ncep.noaa.gov/data/indices/soi, and solar irradiance data from Lean, J. (personal communication 2012).

The Southern Oscillation is an oscillation in the surface air pressure between the tropical eastern and the western Pacific Ocean waters. The SOI only takes into account sea-level pressure. In contrast, the El Niño component of ENSO is specified in terms of changes in the Pacific Ocean sea surface temperature relative to the average temperature. It is considered to be simpler to conduct an analysis in which the temperature is an outcome (dependent variable) without also having (Pacific Ocean) temperature as an input (independent variable). The correlation between SOI and the other ENSO indices is high, so we believe this assumption is robust.

To make it easier to visually assess the relationship between the key climate variables, the data were normalised using statistical Z scores or standardised deviation scores (expressed as "Relative level" in the figures). In a Z-scored data series, each data point is part of an overall data series that sums to a zero mean and variance of 1, enabling comparison of data having different native units. See the individual figure legends for details on the series lengths.

The investigation is conducted using linear regression. SOI and global atmospheric surface temperature are the dependent variables. For these two variables, we tested the relationship between (1) the change in atmospheric $CO_2$ and (2) the variability in its rate of change. We express these $CO_2$-related variables as finite differences, which is a convenient approximation to derivatives (Hazewinkel,2013; Kaufmann et al., 2006). The finite differences used here are of both the first- and second-order types (we label these "first" and "second" differences in the text). Variability is explored using both intra-annual (monthly) data and interannual (yearly) data. The period covered in the figures is shorter than that used in the data preparation because of the loss of some data points due to calculations of differences and of moving averages. The period covered in the figures is shorter than that used in the data preparation because of the loss of some data points due to calculations of differences and of moving averages (in monthly terms of up to 13 x 13), which commenced in January 1960.

Concerning multiple linear regression, Canty *et al.* 38 (2013) note:

> Multiple linear regression of the global surface temperature anomaly has … been used to quantify the relative importance of natural and anthropogenic factors on climate (Lean and Rind (2008, 2009); Foster and Rahmstorf (2011); Kopp and Lean (2011), Zhou and Tung (2013)).

The degree of lead or lag of the variables relative to one another was quantified by means of time-lagged correlations (correlograms). The correlograms were calculated by shifting the series back and forth relative to each other, 1 month at a time. The quantification of the degree of relationship between different plots was carried out using regression analysis to derive the coefficient of determination ($R^2$) for each relationship. Student's t-tests were used to determine the statistical significance of the correlation coefficients.

Smoothing methods are used to the degree needed to produce similar amounts of smoothing for each data series in any given comparison. Notably, to achieve this outcome, series resulting from higher levels of differences require more smoothing. Smoothing is carried out initially by means of a 13-month moving average – this also minimises any remaining seasonal effects. If further smoothing is required, then this is achieved 39 by taking a second moving average of the initial moving average (to produce a double moving average). This is performed by means of a further 13 month moving average, to produce a 13 x 13 moving average.

The rationale of the investigation is to seek the clearest correlations possible between the data series. Because of the reverse seasonality between the hemispheres, it is known that using global averages in climate studies can reduce the clarity of phenomena, even to the extent that different effects can cancel each other out. Hence subsidiary datasets are sometimes used. Such datasets are chosen by studying the correlation between all of the alternative datasets in, say, the dependent variable class, and selecting the one that provides the highest correlation with the independent variable in question. The most important selection to be made here concerns the temperature series, where a number of choices are available. Supplementary Figure 1 shows that, based on measured correlation coefficients, there is more in common between the different temperature series than differences between them. That said, Table 1 (Supplementary Information) shows that for the main candidate influences,

SOI and first-difference $CO_2$, the tropical temperature series fits best. Hence, this series is selected for use in the study.

**Supplementary Figure 1. Comparison of trends between first-difference $CO_2$ and global surface temperature categories:** First-difference atmospheric $CO_2$ (red) and observed monthly mean tropical surface temperature series (HADCRUT4 datasets): northern hemisphere (green); southern hemisphere (blue); tropics (30 degrees N to 30 degrees S latitude) (purple); global (average of northern and southern hemispheres) (black).

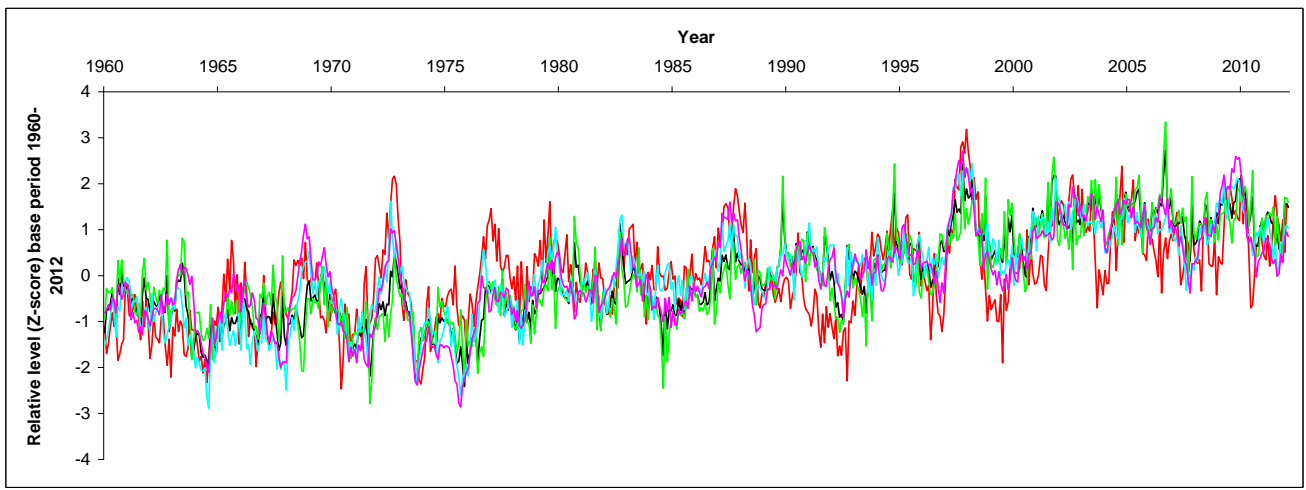

**Supplementary Table 1.** Correlation coefficients (R) from correlogram analysis for first-difference $CO_2$ and SOI against global temperature categories

|  | *13 m ma 1 st diff. atmos. $CO_2$ mlo (season corr)* | *Led by 3 m reverse SOI (KNMI)* |
|---|---|---|
| HadCRUT4.2.0.0 NH | 0.521 | 0.091 |
| HadCRUT4.2.0.0 (NH+SH)/2 | 0.628 | 0.174 |
| HadCRUT4.2.0.0 SH | 0.695 | 0.281 |
| HadCRUT4.2.0.0 Tropics | 0.699 | 0.376 |

**Data smoothing**

Smoothing methods are used to the degree needed to produce similar amounts of smoothing for each data series in any given comparison. Notably, to achieve this outcome, series resulting from higher levels of differences require more smoothing. Smoothing is carried out initially by means of a 13-month moving average – this also minimises any remaining seasonal effects. If further smoothing is required, then this is achieved (Hyndman 2010) by taking a second moving average of the initial moving average (to produce a double moving average). Often, this is performed by means of a further 13 month moving average to produce a 13 x 13 moving average.

**Correlation analysis**

The degree of lead or lag of the variables relative to one another was quantified by means of time-lagged correlations (correlograms). The correlograms were calculated by shifting the series back and forth relative to each other, 1 month at a time.

The quantification of the degree of relationship between different plots was carried out using regression analysis to derive the coefficient of determination ($R^2$) for each relationship. Student's t-tests were used to determine the statistical significance of the correlation coefficients.